\documentclass[aps,prd,a4paper,onecolumn,amsmath,showpacs,superscriptaddress,nofootinbib,preprintnumbers,notitlepage]{revtex4-1}
\usepackage{amssymb,amsmath}
\usepackage{setspace}
\usepackage{graphicx}
\usepackage{color}
\usepackage{dcolumn}
\usepackage{pgfplots}
\usepackage{booktabs}
\usepackage{multirow}
\usepackage{dcolumn}
\usepackage{amsmath}
\usepackage{mathtools}
\usepackage{amsfonts}
\usepackage{amssymb}
\usepackage{epstopdf}
\usepackage{bm}
\usepackage{siunitx}
\usepackage{braket}
\usepackage{enumitem}
\usepackage{soul}
\usepackage{ulem}
\usepackage{color}
\usepackage{transparent}
\usepackage{pifont}
\usepackage[font={small}]{caption}
\newcommand\br{\begin{eqnarray}}
\newcommand\er{\end{eqnarray}}
\newcommand\be{\begin{equation}}
\newcommand\ee{\end{equation}}

\newcommand\bc{\begin{center}}
\newcommand\ec{\end{center}}

\pgfplotsset{compat=1.18}
\begin{document}
\doublespacing
\title {Constructing de Sitter space and  Dark Matter with  Dynamical Tension Strings   \footnote{Honorable mention in the Gravity Research Foundation 2026 Awards for Essays on Gravitation}}
\date{9.02.2025}
\author{Eduardo Guendelman}
\email{eduardoleonguendelman@gmail.com}
\affiliation{Department of Physics, Ben-Gurion University of the Negev, Beer-Sheva, Israel.\\}
\affiliation{Frankfurt Institute for Advanced Studies (FIAS),
Ruth-Moufang-Strasse 1, 60438 Frankfurt am Main, Germany.\\}
\affiliation{Bahamas Advanced Study Institute and Conferences, 
4A Ocean Heights, Hill View Circle, Stella Maris, Long Island, The Bahamas.}
\begin{abstract}
The string tensions can be dynamical in the modified measure formalism and  appear as an additional dynamical degrees of freedom . These tensions may not be universal, instead,  each string generates its own  tension. We then consider a new bulk field that can couple to the strings, the tension  scalar which changes locally the tension along the world sheet. In the case with two string tensions there is a braneworld solution which gives rise to an induced de Sitter space in the brane, avoiding swampland constraints of the standard string theory. Strings with different tension to ours can appear also as Dark Matter and since they share the same space and compactifications as visible matter,  they should lead to Dark copies of the standard model,

\end{abstract}
\maketitle
\section{Dynamical Tension Strings in the Modified Measures Approach}
\label{intro}
String and Brane Theories have been studied as candidates for the theory of all matter and interactions including gravity, in particular string theories \cite{stringtheory}. But string theory has a dimensionful parameter, the tension of the string, in its standard formulation, the same is true for brane theories in their better known formulations, so the appearance of a dimensionful string tension and brane theories from the start appears as somewhat unnatural Previously however, in the framework of a Modified Measure Theory, a formalism originally used for gravity theories, see for example \cite{d,b, Hehl, GKatz, DE, MODDM, Cordero, Hidden}, the tension was derived as an additional degree of freedom \cite{a,c,supermod, cnish, T1, T2, T3}. See also the treatment by Townsend and collaborators \cite{xx,xxx} 
and in \cite{Paul}.

For comparison, a floating cosmological constant is a generic feature of the modified measure theories of gravity \cite{d,b, Hehl, GKatz, DE, MODDM, Cordero, Hidden}, including the covariant formulation of the unimodular theory   
\cite{HT}, which is in fact a particular case of a modified measure theory, as reviewed in \cite{reviewmodmeas} and the tension of the string plays a very similar role to the cosmological constant in four dimensional gravity, but the analogous situation and the role of the cosmological constant is quite different to that of the string tension, because while several world sheets of strings
can exist in the same universe and in this way many strings with different tensions can probe simultaneously the same  region of space time,  the same is not the case for the cosmological constant, where every cosmological constant defines necessarily a different universe.

The standard world sheet string sigma-model action using a world sheet metric is \cite{pol1}, \cite{pol2}, \cite{pol3}

\begin{equation}\label{eq:1}
S_{sigma-model} = -T\int d^2 \sigma \frac12 \sqrt{-\gamma} \gamma^{ab} \partial_a X^{\mu} \partial_b X^{\nu} g_{\mu \nu}.
\end{equation}

Here $\gamma^{ab}$ is the intrinsic Riemannian metric on the 2-dimensional string world sheet and $\gamma = det(\gamma_{ab})$; $g_{\mu \nu}$ denotes the Riemannian metric on the embedding spacetime. $T$ is a string tension, a dimension full scale introduced into the theory by hand. \\

From the variations of the action with respect to $\gamma^{ab}$ and $X^{\mu}$ we get the following equations of motion: $T_{ab} = (\partial_a X^{\mu} \partial_b X^{\nu} - \frac12 \gamma_{ab}\gamma^{cd}\partial_cX^{\mu}\partial_dX^{\nu}) g_{\mu\nu}=0$
and likewise the variation with respect to $X^{\lambda}$ leads to a geodesic eq, for the string.

There are no limitations on employing any other measure of integration different than $\sqrt{-\gamma}$. The only restriction is that it must be a density under arbitrary diffeomorphisms (reparametrizations) on the underlying spacetime manifold.

In the framework of modified-measure string theory theory,  two additional worldsheet scalar fields $\varphi^i (i=1,2)$ are introduced. A new measure density is

\begin{equation}
\Phi(\varphi) = \frac12 \epsilon_{ij}\epsilon^{ab} \partial_a \varphi^i \partial_b \varphi^j.
\end{equation}

Then the modified bosonic string action is (as formulated first in \cite{a} and latter discussed and generalized also in \cite{c})

\begin{equation} \label{eq:5}
S = -\int d^2 \sigma \Phi(\varphi)(\frac12 \gamma^{ab} \partial_a X^{\mu} \partial_b X^{\nu} g_{\mu\nu} - \frac{\epsilon^{ab}}{2\sqrt{-\gamma}}F_{ab}(A)),
\end{equation}

where $F_{ab}$ is the field-strength  of an auxiliary Abelian gauge field $A_a$: $F_{ab} = \partial_a A_b - \partial_b A_a$. \\

It is important to notice that the action (\ref{eq:5}) is invariant under conformal transformations of the intrinsic measure combined with a diffeomorphism of the measure fields: $\gamma_{ab} \rightarrow J\gamma_{ab}$ , $\varphi^i \rightarrow \varphi^{'i}= \varphi^{'i}(\varphi^i)$, such that $\Phi \rightarrow \Phi^{'}= J \Phi$
Here $J$ is the jacobian of the diffeomorphim in the internal measure fields which can be an arbitrary function of the world sheet space time coordinates, so this can be called indeed a local conformal symmetry.
To check that the new action is consistent with the sigma-model one, let us derive the equations of motion of the action (\ref{eq:5}).
The variation with respect to $\varphi^i$ leads to the following equations of motion, $\epsilon^{ab} \partial_b \varphi^i \partial_a (\gamma^{cd} \partial_c X^{\mu} \partial_d X^{\nu} g_{\mu\nu} - \frac{\epsilon^{cd}}{\sqrt{-\gamma}}F_{cd}) = 0.$:
which implies $\gamma^{cd} \partial_c X^{\mu} \partial_d X^{\nu} g_{\mu\nu} - \frac{\epsilon^{cd}}{\sqrt{-\gamma}}F_{cd} = M = const.$
The equations of motion with respect to $\gamma^{ab}$ are

\begin{equation} \label{eq:8}
T_{ab} = \partial_a X^{\mu} \partial_b X^{\nu} g_{\mu\nu} - \frac12 \gamma_{ab} \frac{\epsilon^{cd}}{\sqrt{-\gamma}}F_{cd}=0.
\end{equation}

We see that these equations are the same as in the sigma-model formulation. Namely, taking the trace of (\ref{eq:8}) we get that $M = 0$ and by solving $\frac{\epsilon^{cd}}{\sqrt{-\gamma}}F_{cd}$ from the resulting trace equation  we obtain the conventional Polyakov string equations. \\

A most significant result is obtained by varying the action with respect to $A_a$:

\begin{equation}
\epsilon^{ab} \partial_b (\frac{\Phi(\varphi)}{\sqrt{-\gamma}}) = 0.
\end{equation}

Then by integrating and comparing it with the standard action it is seen that

\begin{equation}
\frac{\Phi(\varphi)}{\sqrt{-\gamma}} = T.
\end{equation}

That is how the string tension $T$ is derived as a world sheet constant of integration opposite to the standard equation (\ref{eq:1}) where the tension is put ad hoc.The variation with respect to $X^{\mu}$ leads to  a geodesic type equation for the motion of the string in the target space. The idea of modifying the measure of integration is proved itself both effective and profitable. This can be generalized to incorporate super symmetry, see for example \cite{c}, \cite{cnish}, \cite{supermod} , \cite{T1}.
For other mechanisms for dynamical string tension generation from added string world sheet fields, see for example \cite{xx} and \cite{xxx}. However the fact that this string tension generation is a world sheet effect 
and not a universal uniform string tension generation effect for all strings has not been sufficiently emphasized before.

\section{Each String  in its own world sheet determines its own  tension. Therefore the  tension may not be universal for all strings}

If we look at a single string, the dynamical string tension theories and the standard string  theories appear indeed indistinguishable, there are however more than one string and/or one brane in the universe then, let us now observe indeed that it does not appear that the string tension or the brane tension derived in the sections above correspond to ¨the¨ string or brane tensions of the theory. The derivation of the string or brane tensions in the previous sections holds for a given string or brane, there is no obstacle that for another string or brane these could acquire a different string or brane  tension.  In other words, the string or brane tension is a world sheet constant, but it does not appear to be a universal constant same for all strings and for all branes.
Similar situation takes place in the dynamical string generation proposed by Townsend for example \cite{xx}, in that paper world sheet fields include an electromagnetic gauge potential. Its equations of motion are those of the Green-Schwarz superstring but with the string tension given by the circulation of the worldsheet electric field around the string. So again, in \cite{xx} also a string will determine a given tension, but another string may determine another tension. 
If the tension is a universal constant valid for all strings, that would require an explanation in the context of these dynamical tension string theories, for example some kind of interactions that tend to equalize string tensions, or that all  strings in the universe originated from the splittings of one primordial string or some other mechanism. 

In any case, if one believes for example in strings , on the light of the dynamical string tension mechanism being a process that takes place at each string independently, we must ask whether all strings have the same string tension and what difference that this makes.

\section{Equations for the Background fields and a new background field}
As discussed by Polchinski for example in \cite{Polchinski} , gravity can be introduced in two different ways in string theory. One way is by recognizing the graviton as one of the fundamental excitations of the string, the other is by considering the effective action of the embedding metric, by integrating out the string degrees of freedom and then the embedding metric and other originally external fields acquire dynamics which is enforced by the requirement of a zero beta function.
These equations fortunately appear to be string tension independent for the critical dimension $D¿26$ in the bosonic string for example, so they will not be changed by introducing different strings with different string tensions, if these tensions are constant along the the world sheet .

However, in addition to the traditional background fields usually considered in conventional string theory, one may consider as well an additional scalar field that induces currents in the string world sheet and since the current couples to the world sheet gauge fields, this produces a dynamical tension controlled by the external scalar field as shown at the classical level in \cite{Ansoldi}. In the next two subsections we will study how this comes about in two steps, first we introduce world sheet currents that couple to the internal gauge fields in Strings and Branes and second we define a coupling to an external scalar field by defining a world sheet currents that couple to the internal gauge fields in Strings and Branes that is induced by such external scalar field. This is very much in accordance to the philosophy of Schwinger \cite{Schwinger} that proposed long time ago that a field theory must be understood by probing it with external sources. 

As we will see however, there will be a fundamental difference between this background field and the more conventional ones (the metric, the dilaton field and the two index anti symmetric tensor field) which are identified with some string excitations as well. Instead, here we will see that a single string does not provide dynamics for this field, but rather when the condition for world sheet conformal invariance is implemented for two strings which sample the same region of space time, so it represents a collective effect instead. 

\subsection{Introducing world sheet currents that couple to the internal gauge fields in Strings and Branes}

If to the action of the string we add a coupling
to a world-sheet current $j ^{a _{2}}$,   i.e. a term
\begin{equation}
    S _{\mathrm{current}}
    =
    \int d ^{2} \sigma
        A _{a _{2}}
        j ^{a _{2}}
    ,
\label{eq:bracuract}
\end{equation}
see   \cite{Escaping},  \cite{summary} , \cite{Life},  \cite{Lidhtlikeandbraneworld} for different applications of this.
 Then the variation of the total action with respect to $A _{a _{2} }$
gives
\begin{equation}
    \epsilon ^{a _{1} a _{2}}
    \partial _{a _{1}} ( \frac{\Phi}{\sqrt{- \gamma}})
     =j ^{a _{2}}
    .
\label{eq:gauvarbracurmodtotact}
\end{equation}

\subsection{Coupling to a bulk scalar field, the tension field}

Suppose that we have an external scalar field $\phi (x ^{\mu})$
defined in the bulk. From this field we can define the induced
conserved world-sheet current
\begin{equation}
    j ^{a_{1}  }
    =
    e \partial _{\mu} \phi
    \frac{\partial X ^{\mu}}{\partial \sigma ^{a}}
    \epsilon ^{a a _{1} }
    \equiv
    e \partial _{a} \phi
    \epsilon ^{a  a _{1}}
    ,
\label{eq:curfroscafie}
\end{equation}
where $e$ is some coupling constant.

Then (\ref{eq:gauvarbracurmodtotact}) can be integrated to obtain
\begin{equation}
  T =  \frac{\Phi}{\sqrt{- \gamma}}
    =
    e \phi + T _{i}
    ,
\label{eq:solgauvarbracurmodtotact2}
\end{equation}
or  equivalently
\begin{equation}
  \Phi
    =
   \sqrt{- \gamma}( e \phi + T _{i})
    ,
\label{eq:solgauvarbracurmodtotact}
\end{equation}

The constant of integration $T _{i}$ may vary from one extended object to the other. Notice that for the string case the interaction is metric independent since the internal gauge field does not transform under the the conformal transformations. This interaction does not therefore spoil the world sheet conformal transformation invariance in the case the field $\phi$ does not transform under this transformation.  One may interpret
(\ref{eq:solgauvarbracurmodtotact} ) as the result of integration out classically (through integration of equations of motion) or quantum mechanically (by functional integration of the internal gauge field, with respect to the boundary condition that characterizes the constant of integration $T _{i}$ for a given string or brane). Then replacing 
$ \Phi
    =
   \sqrt{- \gamma}( e \phi + T _{i})$ back into the remaining terms in the action gives a correct effective action for each string. Each string is going to be quantized with each a different $ T _{i}$. The consequences of an independent quantization of  many strings with different $ T _{i}$

One may interpret 
(\ref{eq:solgauvarbracurmodtotact} ) as the result of integrating out classically (through integration of equations of motion) or quantum mechanically (by functional integration of the internal gauge field, respecting the boundary condition that characterizes the constant of integration  $T _{i}$ for a given string ). Then replacing 
$ \Phi
    =
   \sqrt{- \gamma}( e \phi + T _{i})$ back into the remaining terms in the action gives a correct effective action for each string. Each string is going to be quantized with each one having a different $ T _{i}$. The consequences of an independent quantization of  many strings with different $ T _{i}$
covering the same region of space time will be studied in the next section.    

A similar exercise can be considered for the Target space scale invariance and its spontaneous breaking for the Modified Measure Dynamical Brane Tension Theory. 

\subsection{ The Tension field from World Sheet Quantum Conformal Invariance }

\subsubsection{The case of two different string tensions }

If we have a scalar field coupled to a string or a brane in the way described in the sub section above, i.e. through the current induced by the scalar field in the extended object,  according to eq. 
(\ref{eq:solgauvarbracurmodtotact}), so we have two sources for the variability of the tension when going from one string to the other: one is the integration constant $T _{i}$ which varies from string to string and the other the local value of the scalar field, which produces also variations of the  tension even within the string or brane world sheet. As we discussed in the previous section, we can incorporate the result of the tension as a function of scalar field $\phi$, given as $e\phi+T_i$, for a string with the constant of integration $T_i$ by defining the action that produces the correct 
equations of motion for such string, adding also other background fields, the anti symmetric  two index field $A_{\mu \nu}$ that couples to $\epsilon^{ab}\partial_a X^{\mu} \partial_b X^{\nu}$
and the dilaton field $\varphi $ .
\begin{equation}\label{variablestringtensioneffectiveacton}
S_{i} = -\int d^2 \sigma (e\phi+T_i)\frac12 \sqrt{-\gamma} \gamma^{ab} \partial_a X^{\mu} \partial_b X^{\nu} g_{\mu \nu} + \int d^2 \sigma A_{\mu \nu}\epsilon^{ab}\partial_a X^{\mu} \partial_b X^{\nu}+\int d^2 \sigma \sqrt{-\gamma}\varphi R .
\end{equation}
Notice that if we had just one string, or if all strings will have the same constant of integration $T_i = T_0$. We will take  cases where the dilaton field is a constant or zero, and the antisymmetric two index tensor field is pure gauge or zero, then the demand of conformal invariance for $D=26$ becomes the demand that all the metrics
\begin{equation}\label{tensiondependentmetrics}
g^i_{\mu \nu} =  (e\phi+T_i)g_{\mu \nu}
\end{equation}
will satisfy simultaneously the vacuum Einstein´s equations. 
The interesting case to consider is when there are many strings with different $T_i$, let us consider the simplest case of two strings, labeled $1$ and $2$ with  $T_1 \neq  T_2$ , then we will have two Einstein´s equations, for $g^1_{\mu \nu} =  (e\phi+T_1)g_{\mu \nu}$ and for  $g^2_{\mu \nu} =  (e\phi+T_2)g_{\mu \nu}$, 

\begin{equation}\label{Einstein1}
R_{\mu \nu} (g^1_{\alpha \beta}) = 
  R_{\mu \nu} (g^2_{\alpha \beta}) = 0
\end{equation}

These two simultaneous conditions above impose a constraint on the tension field
 $\phi$, because the metrics $g^1_{\alpha \beta}$ and $g^2_{\alpha \beta}$ are conformally related, but Einstein´s equations are not conformally invariant, so the condition that Einstein´s equations hold  for both  $g^1_{\alpha \beta}$ and $g^2_{\alpha \beta}$
is highly non trivial. Then for these situations, we have,

\begin{equation}\label{relationbetweentensions}
e\phi+T_1 = \Omega^2(e\phi+T_2)
\end{equation}
 which leads to a solution for $e\phi$
 
\begin{equation}\label{solutionforphi}
e\phi  = \frac{\Omega^2T_2 -T_1}{1 - \Omega^2} 
\end{equation}
which leads to the tensions of the different strings to be
\begin{equation}\label{stringtension1}
 e\phi+T_1 = \frac{\Omega^2(T_2 -T_1)}{1 - \Omega^2} \\\\\  and,  \\\\
 e\phi+T_2 = \frac{(T_2 -T_1)}{1 - \Omega^2} 
\end{equation}

Both tensions can be taken as positive if $T_2 -T_1$ is positive and $\Omega^2$ is also positive and less than $1$.

\section{Braneworld Solutions Leading to de Sitter Space}

 We have  seen in previous papers how the Tension field gets dynamics in the case of two strings with different tensions, in Refs    \cite{Escaping},  \cite{summary} , \cite{Life},  \cite{Lidhtlikeandbraneworld} and 
\cite{Targetspacemodifiedmeasure}, where it was shown the dynamical tension theories can be produce braneworld scenarion and other interesting effects like avoiding the swampland constraints. 
\subsubsection{Flat space in Minkowski coordinates and flat space after a special conformal transformation }

The flat spacetime in Minkowski coordinates is,

 \begin{equation}\label{Minkowski}
 ds_1^2 = \eta_{\alpha \beta} dx^{\alpha} dx^{\beta}
\end{equation}

where $ \eta_{\alpha \beta}$ is the standard Minkowski metric, with 
$ \eta_{00}= 1$, $ \eta_{0i}= 0 $ and $ \eta_{ij}= - \delta_{ij}$.
This is of course a solution of the vacuum Einstein´s equations.

We now consider the conformally transformed metric

 \begin{equation}\label{Conformally transformed Minkowski}
 ds_2^2 = \Omega(x)^2  \eta_{\alpha \beta} dx^{\alpha} dx^{\beta}
\end{equation}
where conformal factor coincides with that obtained from the special conformal transformation
\begin{equation}\label{ special conformal transformation}
x\prime ^{\mu} =  \frac{(x ^{\mu} +a ^{\mu} x^2)}{(1 +2 a_{\nu}x^{\nu} +   a^2 x^2)}
 \end{equation}
for a certain D vector $a_{\nu}$.  which gives $\Omega^2 =\frac{1}{( 1 +2 a_{\mu}x^{\mu} +   a^2 x^2)^2} $
In summary, we have two solutions for the Einstein´s equations,
 $g^1_{\alpha \beta}=\eta_{\alpha \beta}$ and 
 
 \begin{equation}\label{ conformally transformed metric}
 g^2_{\alpha \beta}= \Omega^2\eta_{\alpha \beta} =\frac{1}{( 1 +2 a_{\mu}x^{\mu} +   a^2 x^2)^2} \eta_{\alpha \beta}
 \end{equation}
 
 We can then study the evolution of the tensions using 
 $\Omega^2 =\frac{1}{( 1 +2 a_{\mu}x^{\mu} +  a^2 x^2)^2}$.
 We will consider the cases where  $a^2 \neq 0 $.

  We now consider the case when $a^\mu$ is not light like and we will find that for $a^2 \neq 0$, irrespective of sign, i.e. irrespective of whether  $a^\mu$ is space like or time like, we will have thick  Braneworlds  where strings can be constrained  between two concentric spherically symmetric bouncing higher dimensional spheres and where the distance between these two  concentric spherically symmetric bouncing higher dimensional spheres approaches zero at large times.
  The string tensions of the strings one and two are given by
    \begin{equation}\label{stringtension1forBraneworld}
 e\phi+T_1 = \frac{(T_2-T_1)( 1 +2 a_{\mu}x^{\mu} +  a^2 x^2)^2}{( 1 +2 a_{\mu}x^{\mu} +  a^2 x^2)^2-1}=  \frac{(T_2-T_1)( 1 +2 a_{\mu}x^{\mu} +  a^2 x^2)^2}{(2 a_{\mu}x^{\mu} +  a^2 x^2)(2+2 a_{\mu}x^{\mu} +  a^2 x^2)}
\end{equation}
  \begin{equation}\label{stringtension2forBraneworld}
 e\phi+T_2 = \frac{(T_2-T_1)}{( 1 +2 a_{\mu}x^{\mu} +  a^2 x^2)^2-1}=  \frac{(T_2-T_1)}{(2 a_{\mu}x^{\mu} +  a^2 x^2)(2+2 a_{\mu}x^{\mu} +  a^2 x^2)}
\end{equation}
  In fact it is the case that the string tensions can only change sign by going first to infinity and then come back from minus infinity. We can now recognize at those large times the locations where the string tensions go to infinity, which  are determined by the conditions $2 a_{\mu}x^{\mu} +  a^2 x^2 = 0$
or $2 +2 a_{\mu}x^{\mu} +  a^2 x^2 = 0$ 

Let us start by considering the case where  $a^\mu$ is time like, then without loss of generality,  we can take  $a^\mu = (A, 0, 0,...,0)$.

The condition $2 a_{\mu}x^{\mu} +  a^2 x^2 = 0$, if $A \neq 0$ implies then that
\begin{equation}\label{bubbleboundaryforBraneworld1a}
 x^2_1  + x^2_2 + x^2_3.....+ x^2_{D-1}- (t+ \frac{1}{A})^2 = -\frac{1}{A^2}
\end{equation}

The other boundary of infinite string tensions,  is given by,
\begin{equation}\label{bubbleboundaryforBraneworld1b}
 x^2_1  + x^2_2 + x^2_3.....+ x^2_{D-1}- (t+ \frac{1}{A})^2 = \frac{1}{A^2}
\end{equation}

We see that $2 +2 a_{\mu}x^{\mu} +  a^2 x^2 = 0$  represents an exterior boundary which has a bouncing  motion with a minimum radius $\frac{1}{A}$ at $t = - \frac{1}{A}$ , 
The denominator tensions are positive between these two bubbles.
So for $T_2 -T_1$ positive the tensions are positive and diverge at the boundaries defined above.

For large positive or negative times, the difference between the upper radius  and the lower radius goes to zero as  $t \rightarrow \infty$
since $\sqrt{\frac{1}{A^2} +(t+ \frac{1}{A})^2 } -\sqrt{-\frac{1}{A^2} +(t+ \frac{1}{A})^2 }\rightarrow \frac{1}{t A^2}\rightarrow 0 $
of course the same holds  $t \rightarrow -\infty$.
This means that for very large early or late times the segment where the strings would be confined (since they will avoid having infinite tension) will be very narrow and the resulting scenario will be that of a brane world for late or early times, while in the bouncing region the inner surface does not exist.
Notice that this kind of braneworld scenario is very different to the ones previously studied (\cite{Rubakov} - \cite{KK}), in particular both gravity (closed strings) and gauge fields (open strings) are treated on the same footing, since the mechanism that confines the strings between the two surfaces relies only on the string tension becoming very big,not on what type of string we are dealing with.

Notice that only (\ref{bubbleboundaryforBraneworld1b})  can represent a physical surface where strings can live and condensate, but not 
 (\ref{bubbleboundaryforBraneworld1a})  where no strings can live or reach, since this hypersurface represents superluminal motion, but  (\ref{bubbleboundaryforBraneworld1b})  represents a de Sitter space in fact, which is constructed as the juncture of two flat spaces and this construction has been studied without reference to string theory in 
\cite{guendelmaportnoy}. 
defining  $r^2=x^2_1  + x^2_2 + x^2_3$, say in a three dimensional embedding space, where $ r^2 = (t+ \frac{1}{A})^2 + \frac{1}{A^2}$, which implies that $-dt^2 + dr^2 = -dt^2 \frac{1}{(A^2((t + \frac{1}{A})^2 +\frac{1}{A^2})}  = -  d\tau^2$, 
which defines the proper time observed by a co-moving observer (that is an observer with $\theta$ and $\phi$  constant) in the bubble as a function of the embedding time. The induced space and induced metric in the bubble is then perfectly well defined from either side due to the fact that the conformal factor at the junction of the two spaces is one. 
The  above equation relating $t$ and  $\tau$ can of course be integrated, giving 
$ At +1 = sinsh (A\tau) = $, and we obtain $ r^2 = \frac{1}{A^2}  cosh^2 (A\tau)$, 
so, altogether the induced metric is a manifestly de Sitter $2+1$ metric
$$ds^2 = d\tau^2 + \frac{1}{A^2}  cosh^2 (A\tau) ( d\phi^2 + sin(\theta)^2  d\theta^2 ) $$
a very well known representation of a $2+1$ de Sitter space. The generalization to higher dimensions is straightforward so the metric induced on (\ref{bubbleboundaryforBraneworld1b}) is always a de Sitter space. In summary, the construction of a de Sitter space is possible in a dynamical tension braneworld scenario, in contrast to the standard string theory, where the swampland constraints \cite{swampland} forbid this space from existing as a stable solution. 

\subsection{Target space scale invariance and restoration providing further stabilization of the de Sitter solution } 

The ordinary string theory does not have  target space scale invariance, which is very much related to the fact
that there is a definite scale in the theory, the string tension. Indeed, in the ordinary string theory, a scale transformation of the embeddng metric $g_{\mu \nu} $, of the form, 
$g_{\mu \nu}  \rightarrow   \omega g_{mu \nu}$
where $\omega $ is a constant, 
is not part of a symmetry of the Polyakov action, but in the dynamical tension string theory, this transformation is a part of a symmetry provided the world sheet gauge fields transforms as 
 $A_{a}  \rightarrow   \omega A_{a}$ , the  measure transforms as
 $\Phi(\varphi) \rightarrow   \omega ^{-1} \Phi(\varphi)  $
 and the tension field transforms as,
 $\phi \rightarrow   \omega ^{-1} \phi $
 The integration of the equations of motions leads to the spontaneous generation of the string tension and at the same time, the spontaneous generation of breaking of the target space global scale invariance. At the locations where the string tension becomes infinite, we find restoration of the target space scale invariance. Such scale symmetric state must be reinforced by the functional integration of the symmetry variables corresponding to performing a target space scale transformation, which should reinforce the stability of the de Sitter solution .  
 \section{Appearence of a new type of brane. Comparison with D branes.}
 There is a fundamental difference between the branes studied here and D branes for example, which is that the D branes are surfaces where open strings end . The D branes constrain open strings to end at these branes, while closed strings are unconstrained and can wonder into the bulk.
In contrast, the branes that are considered here are generated by the space- time dependence of the dynamical tension, that pushes all strings, whether open or closed into the induced de Sitter 
Hyper space Surface (\ref{bubbleboundaryforBraneworld1b}). So for example both standard model particles, including quarks, leptons, gauge fields and Englert - Higgs bosons  as well as gravity waves will all be constrained to this type  of branes.

   \section{Disappearance of standard string interactions between strings with different tensions and Emergence of a new model for Dark Matter}
  We have see a new type of interactions between strings, in fact between string with different tensions, mediated  by the tension field, but at the same time, the standard interactions of strings disappear. This is because these standard interactions have been formulated only for strings with the same tension, so these kind of interactions, disappear now, since they consist of splitting or joining, etc. of strings  which only make sense for strings with the same tension. Even a simpler model that does not involve a tension field, would show similar effect. The idea that strings with a different tension could explain the dark matter was formulated in 
\cite{stringtheorywithadifftensionasDM}. 
  The tension field should be necessary to show some interaction between strings with different tensions, although not the standard string  interactions.  Also the cosmological emergence of different tension strings should be explained.
  One should point out however that strings with all types of tensions contribute to the structure of space time. The space time string tension metrics are related by a conformal transformation, so , in this way the effects of one string type affects the common metric and this back reacts on the other space time metric. So both strings species gravitate.  We then look at what we know about our universe. There is indeed a big sector of our universe that does not share standard model interactions with us, the dark sector.  But in the context of our findings here, we see that Dark matter to us may consist of matter made out of strings with different tensions because of the decoupling of standard string interactions for strings with different tensions. 
  It seems that a framework for implementing a dark matter scenario that mimics the standard model, but with different parameters could be based on \cite{stringtheorywithadifftensionasDM} as we have pointed  in \cite{stringtheorywithadifftensionasDMADDEDUM}. 
 This is because all strings with our and other tensions share the same space time, and in particular,  they should share the same compactification and since it is the compactification is what determines the patterns of the particle spectrum,  so the Dark Strings are likely to organize themselves in a similar way as the visible matter, leading probably to copies of the standard model with different parameters, since the string tension is not the same. This should be somewhat related to the models of dark matter that mimic the standard model,  as for example in references \cite{DARKMATTERMIMIKINGSTANDARDMODEL} , \cite{Randall} , \cite{Vikman}.


\end{document}